# Productively Expressing High-performance Spatial Designs of Givens Rotation-based QR Decomposition Algorithm

T2S Working Note #1

Hongbo Rong
Parallel Computing Lab (PCL), Intel Corporation
hongbo.rong@intel.com

**Abstract**

QR decomposition is used prevalently in wireless communication. In this paper, we express the Givens-rotation-based QR decomposition algorithm on a spatial architecture using T2S (Temporal To Spatial), a high-productivity spatial programming methodology for expressing high-performance spatial designs.

There are interesting challenges: the loop iteration space is not rectangular, and it is not obvious how the imperative algorithm can be expressed in a functional notation – the starting point of T2S.

Using QR decomposition as an example, this paper elucidates some general principle, and de-mystifies high-performance spatial programming.

The paper also serves as a tutorial of spatial programming for programmers who are not mathematicians, not expert programmers, and not experts on spatial architectures, but still hope to intuitively identify a high-performance design and map to spatial architectures efficiently.

## 1 Introduction

Given a matrix $A_{MN}$, QR decomposition finds matrix $Q_{MM}$ and $R_{MN}$ such that
(1) $A=QR$,
(2) $QQ^H=I$, and
(3) $R$ is upper-triangular.

Here $Q^H$ is the conjugate transpose of $Q$, and $I$ is an identity matrix.

We consider the Givens-rotation-based QR decomposition algorithm [1, 2] for a spatial architecture like Field-Programmable Gate Array (FPGA) [3] or Coarse-Grain Reconfigurable Architecture (CGRA) [4].

We would like to express the algorithm in T2S (Temporal To Spatial) [5], a high-productivity spatial programming methodology for expressing high-performance spatial designs. A T2S program is a specification, which includes a *temporal definition* and a *spatial mapping*. The temporal definition is a functional notation of the algorithm, and the spatial mapping describes how to map the algorithm to a spatial architecture.

The Givens-rotation-based QR decomposition presents some interesting challenges to the above T2S methodology: the loop iteration space is not rectangular, and it is not obvious how the imperative algorithm can be expressed in a functional notation – the starting point of T2S.

In addressing the challenges, we find the following process:
1. Building a dataflow graph.
   A node in a dataflow graph is a loop iteration around an operation in a given (imperative) algorithm. To coarsen the granularity, several operations at the same loop level can be grouped into the same node.
2. Expressing each type of nodes as a dataflow function.
   For each type of nodes, identify its dataflow patterns. Based on the patterns, one can express the type of nodes in a dataflow function.
3. Scheduling the functions.
   For example, specify the communication between the functions, unroll some loops, drain the results, etc.

The above process should be universally applicable to many problems, not limited to QR decomposition.

## 2 QR decomposition based on Givens rotation

Solve this equation:
(4) $Ay=z$,
where y and z are column vectors. A and z are known, and y is unknown.

Since $A=QR$, multiply $Q^H$ to both sides of equation (4), and we get
(5) $Ry=Q^Hz$.

Now that R is upper-triangular, y can be easily solved with back substitution in a subsequent phase.

When automatically transforming the above equation (4) to equation (5), for convenience, A and z are combined together into a new matrix A', and then $Q^H$ is multiplied to A':

(6)  A' = (A | z)
(7)  $Q^H$A' = ($Q^H$A | $Q^H$z) = (R | $Q^H$z)

$Q^H$ is calculated as the product of a series of Givens rotation matrices. Let us illustrate the process with an example. Say, M=N=4. Then

$Q^H = G_{4,3}G_{3,2}G_{4,2}G_{2,1}G_{3,1}G_{4,1}$,

where every $G_{row, col}$ is a Givens rotation matrix, which when multiplied with A', zeros $A'_{row, col}$, and updates the other elements that are in the same column or the right columns, and in the same row and the row above. See the illustration below:

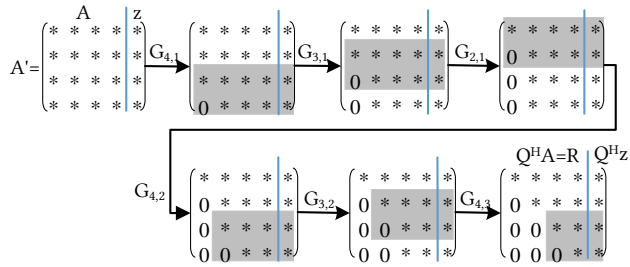

Here a shaded box shows the elements that are changed due to a Givens rotation.

Each $G_{row, col}$ is determined by two parameters $c_{row, col}$ and $s_{row, col}$. For the above example,

$$G_{4,1} = \begin{bmatrix} 1 & 0 & 0 & 0 \\ 0 & 1 & 0 & 0 \\ 0 & 0 & c_{4,1} & s_{4,1} \\ 0 & 0 & -s_{4,1} & c_{4,1} \end{bmatrix},$$

and so on.

Figure 1 shows the QR decomposition algorithm with Givens rotation, where f and g are two functions[1].

Finally, notice that after multiplying $Q^H$ to A', A has been changed to an upper triangle R, and z has been changed to another column vector $Q^H$z. Since z has been transformed in the same process, we do not really need to calculate $Q^H$ explicitly.

## 2.1  Building a dataflow graph

*The key to map an algorithm to a spatial design is to understand the dataflow.* To figure out the dataflow, we wrote a trivial C program for the algorithm in Figure 1, and instrumented the program to print out the data accessed in each iteration. For our M=N=4 example, the results are shown in Figure 2. From the results, it is straightforward to draw a dataflow graph (Figure 3), where *every node is a loop iteration, and data items flow between the nodes.*

There are two type of nodes in the dataflow graph: round and rectangular nodes. The rounds nodes correspond to the computation in Line 3 of Figure 1, which outputs 4 values: $c_{row, col}$, $s_{row, col}$, $A'_{row, col}$, and $A'_{row-1, col}$, represented in different-colored output lines (although in Figure 3, for simplicity, we have merged $c_{row, col}$ and $s_{row, col}$ and represent them in the same line). Similarly, the rectangular nodes correspond to the computation in Line 5 of Figure 1.

## 2.2  Expressing each type of nodes as a dataflow function

There are two type of nodes in the dataflow graph in Figure 3: round nodes and rectangular nodes. Correspondingly, we need define two functions: X and Y, for the round and rectangular nodes, respectively.

(1) The first function, X, is for the computation in Line 3 of Figure 1, which is within two loops (col and row), and returns 4 values. Therefore, we define the function as X(col, row) and let it return a tuple of 4 values. The 4 values in a returned tuple will be referred to as X(col, row)[i], where i=0, 1, 2, 3.

To define the function, we identify all its dataflow patterns from the dataflow graph. There are 4 patterns identified, as exemplified in Figure 4. The first 3 patterns appear at certain boundaries of the graph, and the last one is the basic pattern.

(a) Both inputs are from memory.
 This happens only at the top row of the dataflow graph, i.e. only when col=1 and row=M. Therefore,
 X(1, M) = f($A'_{M, 1}$, $A'_{M-1, 1}$)
(b) One input is from X, the other from memory.
 This happens only when col=1 but row≠M.
 X(1, row) = f(X(1, row+1)[3],
   $A'_{row-1, 1}$).
 Note that X(1, row+1)[3] refers to the 4th returned value of X(1, row+1), as we start counting from 0.
(c) Both inputs are from Y.
 This happens only when col≠1 but row=M.
 X(col, M) = f(Y(col-1, M, col)[0],
   Y(col-1, M-1, col)[0]).
(d) One input is from X, the other from Y.
 X(col, row) = f(X(col, row+1)[3],
   Y(col-1, row-1, col)[0]).

---

[1] Function f calculates $c_{row,col} = \frac{x}{\sqrt{x^2+y^2}}$ and $s_{row,col} = -\frac{y}{\sqrt{x^2+y^2}}$, where $x = A'_{row-1,col}$ and $y = A'_{row,col}$ [2]. The two parameters completely determines a Givens rotation matrix $G_{row, col}$. Then function f and g multiply $G_{row, col}$ with column col and col+1~N+1 of matrix A', respectively.

**Input:** A' = ( A | z)   // Extend A with z
**Output:** $Q^H A' = (Q^H A \mid Q^H z)$   // Transform both A and z simultaneously
**Algorithm:**
1  for col = 1 to N
2     for row = M downto col+1
     // Compute c and s with the two elements in row and row -1 in this column. And update
     // the two elements.
3       $c_{row, col}, s_{row, col}, A'_{row, col}, A'_{row-1, col} = f(A'_{row, col}, A'_{row-1, col})$
     // Use c and s to update the other elements in the two rows that are to the right of the column.
4       for k = col + 1 to N + 1
5         $A'_{row, k}, A'_{row-1, k} = g(c_{row, col}, s_{row, col}, A'_{row, k}, A'_{row-1, k})$

**Figure 1. The QR decomposition algorithm based on Givens rotation.**

```
Iteration    Data read
(col,row,k)  and written
    1,4,-    c,s[4,1](W) A'[4,1] A'[3,1]            1,2,4    c,s[2,1](R) A'[2,4] A'[1,4]
    1,4,2    c,s[4,1](R) A'[4,2] A'[3,2]            1,2,5    c,s[2,1](R) A'[2,5] A'[1,5]
    1,4,3    c,s[4,1](R) A'[4,3] A'[3,3]            2,4,-    c,s[4,2](W) A'[4,2] A'[3,2]
    1,4,4    c,s[4,1](R) A'[4,4] A'[3,4]            2,4,3    c,s[4,2](R) A'[4,3] A'[3,3]
    1,4,5    c,s[4,1](R) A'[4,5] A'[3,5]            2,4,4    c,s[4,2](R) A'[4,4] A'[3,4]
    1,3,-    c,s[3,1](W) A'[3,1] A'[2,1]            2,4,5    c,s[4,2](R) A'[4,5] A'[3,5]
    1,3,2    c,s[3,1](R) A'[3,2] A'[2,2]            2,3,-    c,s[3,2](W) A'[3,2] A'[2,2]
    1,3,3    c,s[3,1](R) A'[3,3] A'[2,3]            2,3,3    c,s[3,2](R) A'[3,3] A'[2,3]
    1,3,4    c,s[3,1](R) A'[3,4] A'[2,4]            2,3,4    c,s[3,2](R) A'[3,4] A'[2,4]
    1,3,5    c,s[3,1](R) A'[3,5] A'[2,5]            2,3,5    c,s[3,2](R) A'[3,5] A'[2,5]
    1,2,-    c,s[2,1](W) A'[2,1] A'[1,1]            3,4,-    c,s[4,3](W) A'[4,3] A'[3,3]
    1,2,2    c,s[2,1](R) A'[2,2] A'[1,2]  continue  3,4,4    c,s[4,3](R) A'[4,4] A'[3,4]
    1,2,3    c,s[2,1](R) A'[2,3] A'[1,3]            3,4,5    c,s[4,3](R) A'[4,5] A'[3,5]
```

**Figure 2. The iterations and the data read and written by them, assuming M=N=4. R/W: written/read. A data item is both read and written if not annotated.**

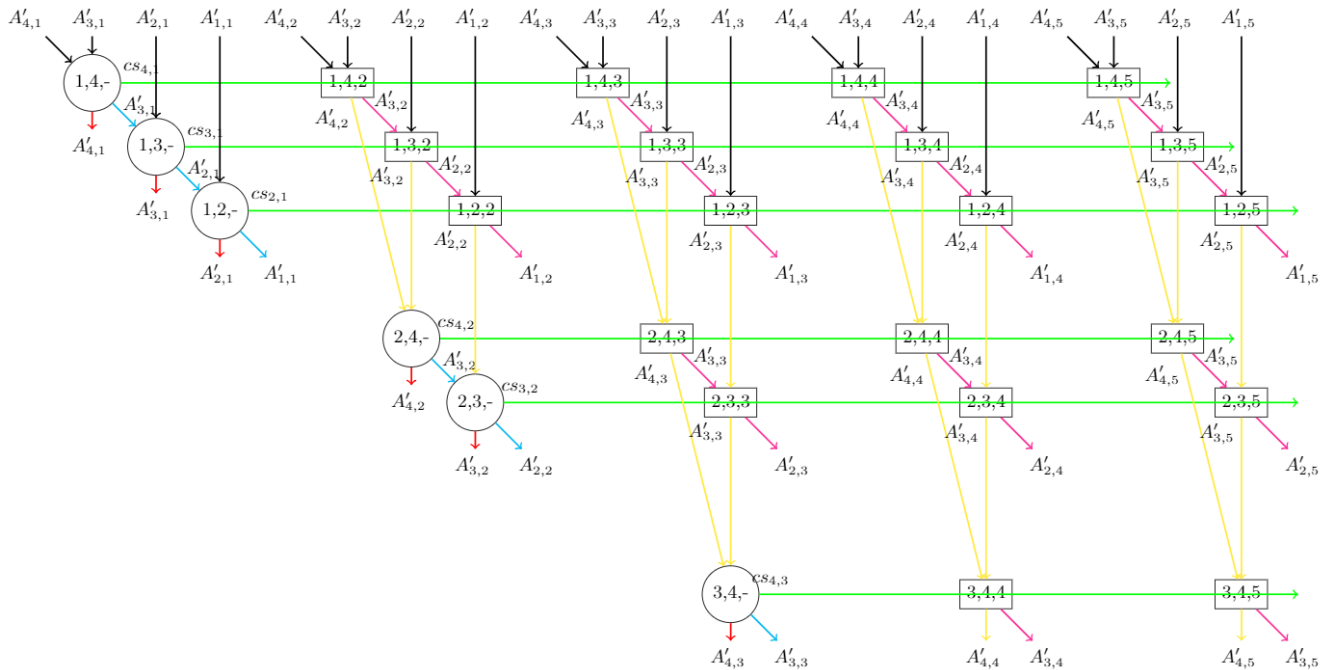

**Figure 3. Dataflow graph.** Here $cs_{4,1}$ means $c_{4,1}$ and $s_{4,1}$, and so on.

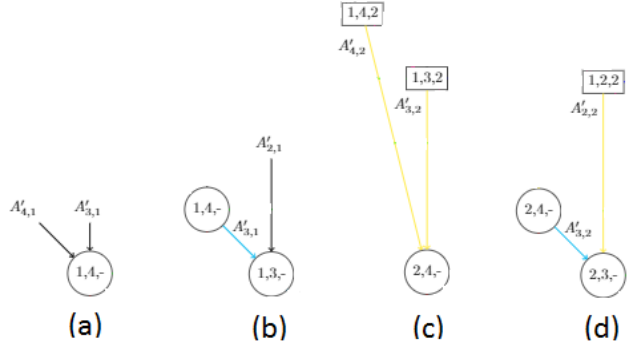

Figure 4. Four cases of function X.

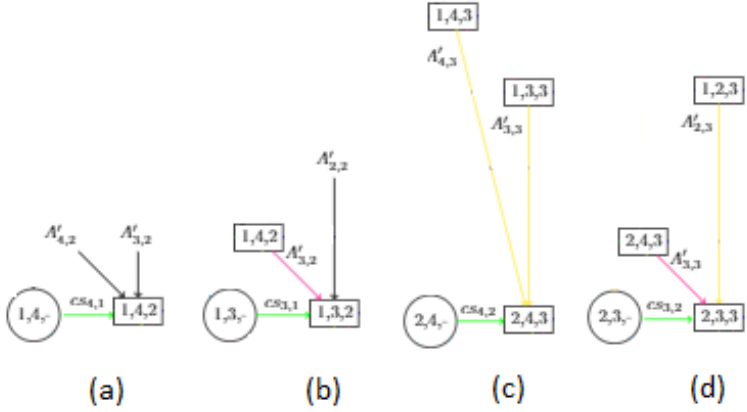

Figure 5. Four cases of function Y.

(2) The second function, Y, is for the computation in Line 5 of Figure 1, which is within 3 loops (col, row, and k), and returns 4 values. Therefore, we define the function as Y(col, row, k) and let it return a tuple of 2 values.

To define the function, we identify all its dataflow patterns from the dataflow graph.

There are 4 patterns identified for function Y, as shown in Figure 5. In these patterns, except the c and s input are from an X node, the other 2 inputs for A' have 4 different cases:
(a) Both are from memory.
(b) One is from Y, the other from memory.
(c) Both are from Y of the previous column.
(d) One is from Y of this column, but the other is from Y of the previous column.

One can write the definitions for Y in exactly the same way as we did before for function X. We skip the details here.

Figure 6 shows the T2S specification. Line 1 says that there is one input A', which is a 2-dimensional floating-point array. And Line 2 says that M and N are symbolic constants.

Line 3~14 define the two functions X and Y. Among them, Line 5~8 defines function X in the formulas we found before. Line 9 sets the bounds of the loops in the function. Similarly, Line 10~13 defines function Y, and Line 14 sets up the loop bounds.

## 2.3 Scheduling the functions

Line 15~16 make X and Y connect with channels. For example, Line 15 says that for X, both the X and Y inputs should come from channels.

Assume there are sufficient hardware resources on the spatial architecture so that we can directly map the entire dataflow graph to the architecture. Line 17~18 fully unroll the iteration spaces of the two functions. As we said, the functions have been made to communicate with channels.

After unrolling, each iteration becomes a hardware PE (Processing Element) on the spatial architecture. Line 19 further says that across a row of PEs of function Y, the c and s  values   should   be

```
1   Parameter A'(float, 2);
2   Assumption symbolic_constants(M, N);

// Define the two functions:
3   Func X, Y;
4   Var   col, row, k;
5   X(1, M) = f(A'_{M, 1}, A'_{M-1, 1});
6   X(1, row) = f(X(1, row+1)[3], A'_{row-1, 1});
7   X(col, M) = f(Y(col-1, M, col)[0], Y(col-1, M-1, col)[0]);
8   X(col, row) = f(X(col, row+1)[3], Y(col-1, row-1, col)[0]);
9   X.bound(col, 1:1:N).bound(row, M:-1:col+1);
10  Y(1, M, k) = f(X(1, M)[0], X(1, M)[1], A'_{M, k}, A'_{M-1, k});
11  Y(1, row, k) = f(X(1, row)[0], X(1, row)[1], Y(1, row+1, k)[1], A'_{row-1, k});
12  Y(col, M, k) = f(X(col, M)[0], X(col, M)[1], Y(col-1, M, k)[0], Y(col-1, M-1, k)[0]);
13  Y(col, row, k) = f(X(col, row)[0], X(col, row)[1], Y(col, row+1, k)[1], Y(col-1, row-1, k)[0]);
14  Y.bound(col, 1:1:N).bound(row, M:-1:col+1).bound(k, col+1:1:N +1);

// Make X and Y connect via channels:
15  X.channel(X, Y);
16  Y.channel(X, Y);

// Fully unroll the iteration spaces:
17  X.unroll(col).unroll(row);
18  Y.unroll(col).unroll(row).unroll(k);

// Relay c and s through Y horizontally:
19  Y.relay(X, {0, 1}, <0, 0, 1>);

// Conditionally save the outputs as the final results:
20  X.store({3}, row ==col+1);
21  Y.store({1}, row==col+1);
22  Y.store({0}, row==col+1 && row == M);
```

**Figure 6. Expressing the algorithm.**

| F.bound(var, range) | Set the bound of a loop variable in Func F to be the given range. The range is in the format of lower bound : step : upper bound, where lower and upper bound are included in the range. |
|---|---|
| F.channel(F1, F2, ...) | For a calling of Func F1 (or F2, ...) in Func F, the value is returned via a channel. |
| F.relay(F1, vector)<br>F.relay(F1, indices, vector) | In an unrolled iteration space of F, determined by some F.unroll(..) before, forward the returned value of a calling of Func F1 along the direction of the given vector. If F1 returns a tuple, indices can be given to choose only some elements of the returned value. |
| F.store(condition)<br>F.store(indices, condition) | Store the results of Func F back to the external memory only under the given condition. If F produces a tuple, indices can be given to choose only some elements of the value. |

**Table 1. Directives.**

forwarded from a PE to the next PE, as shown by the green lines in Figure 3.

Finally, Line 20~22 sets the conditions to output the final results.

Some directives used in the T2S specification are described in Table 1.

The specification in Figure 6 fully unrolls the iteration spaces of the two functions, which maps the dataflow graph shown in Figure 3 directly to the spatial hardware. Of course, the specification is not limited to M=N=4 as shown in Figure 3, since M and N are symbolic constants in the specification.

## 2.4 A variation of the design

If hardware resources on FPGA is limited, one may slightly change the specification by not unrolling some loop(s). For example, if we do not specify unrolling the row loop for function X and Y, all the iterations like X(col, *) or Y(col, *, k), where * means any row, will be mapped to the same X or Y PE, and thus the specification will result in a hardware design as shown in Figure 7, which is exactly the design proposed in literatures [1, 2].

## 3 Conclusion

We have explored how QR decomposition based on Givens rotation can be expressed in T2S style. The imperative algorithm is converted to a functional notation by building a dataflow graph and identifying patterns from the graph. Then it is straightforward to express the mapping of the functions to a spatial architecture.

We are working on an implementation of a T2S compiler, including the new language features shown in this document.

## References

1   Altera Application Note 506. QR Matrix Decomposition. Feb. 2008, version 2.0.
2   Wang, X. and Leeser, M. 2009. A truly two-dimensional systolic array FPGA implementation of QR decomposition. ACM Trans. Embedd. Comput. Syst. 9, 1, Article 3 (October 2009), 17 pages.
3   Wikipedia. Field-programmable gate array. https://en.wikipedia.org/wiki/Field-programmable_gate_array
4   Vaishali Tehre and Ravindra Kshirsagar. Survey on Coarse Grained Reconfigurable Architectures. *International Journal of Computer Applications* 48(16):1-7, June 2012.
5   Hongbo Rong. Programmatic Control of a Compiler for Generating High-performance Spatial Hardware. https://arxiv.org/abs/1711.07606.

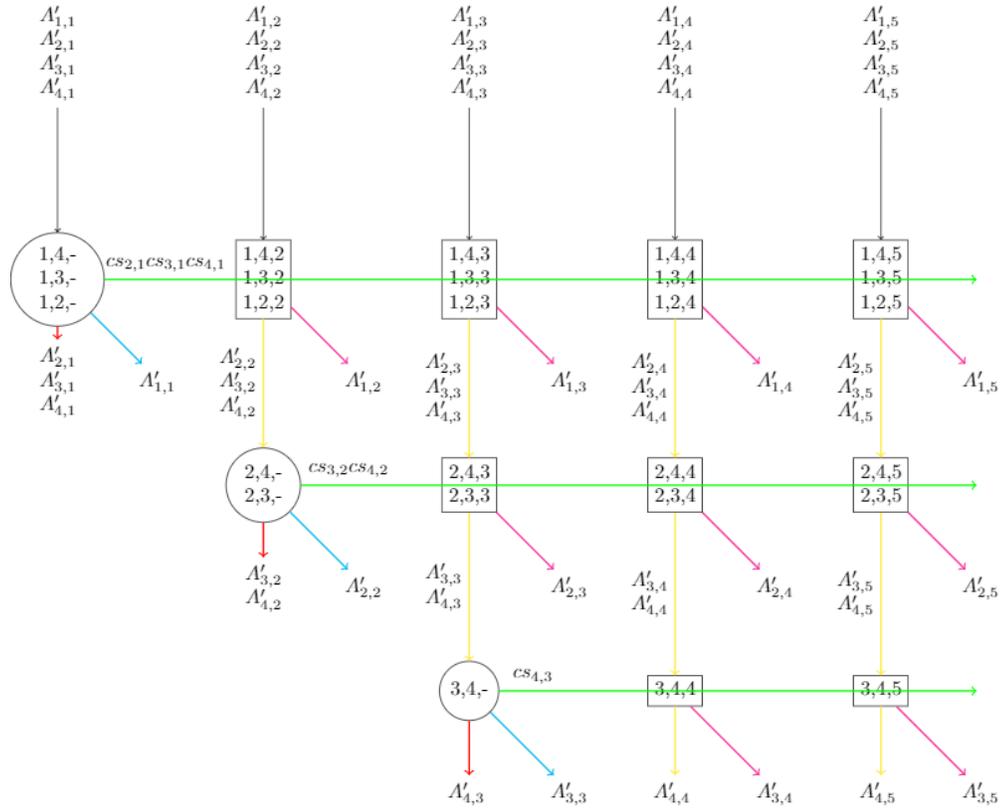

Figure 7. The hardware design if not unrolling the row loop.